\documentclass[a4paper,11pt]{article}
\usepackage{a4wide,amsmath,amssymb,bbm}
\usepackage[english]{babel}

\pdfoutput=1

\newcommand{\bea}{\begin{eqnarray}}
\newcommand{\eea}{\end{eqnarray}}

\newcommand{\nn}{\nonumber}

\newcommand{\ket}[1]{|{#1}\rangle}

\usepackage{graphicx}
\usepackage{feynmp}
\usepackage{color}

\usepackage[nosort]{cite}

\def\unit{\protect{{1 \kern-.28em {\rm l}}}}

\DeclareGraphicsExtensions{.pdf}  

\makeatletter

\@addtoreset{equation}{section}\makeatother

\begin{document}

\newsavebox{\feynmanrules}
\sbox{\feynmanrules}{
\begin{fmffile}{diagrams} 


\fmfset{thin}{0.6pt}  
\fmfset{dash_len}{4pt}
\fmfset{dot_size}{1thick}
\fmfset{arrow_len}{6pt} 



\begin{fmfgraph}(80,40)
\fmfkeep{sunset}
\fmfleft{i}
\fmfright{o}
\fmf{plain,tension=5}{i,v1}
\fmf{plain,tension=5}{v2,o}
\fmf{plain,left,tension=0.4}{v1,v2,v1}
\fmf{plain}{v1,v2}
\fmfdot{v1,v2}
\end{fmfgraph}

\begin{fmfgraph*}(120,75)
\fmfkeep{doubblebubble}
    \fmfleft{i1,i2,i3}
    \fmfright{o1,o2,o3}
    \fmf{plain}{i1,v1,o1}
    \fmffreeze
    \fmf{phantom}{i2,v2,o2}             
    \fmf{phantom}{i3,v3,o3}
    \fmf{plain,left}{v1,v2,v1}
    \fmf{plain,left}{v2,v3,v2}
    \fmfdot{v1,v2}
\end{fmfgraph*}

\begin{fmfgraph*}(72,25)
\fmfkeep{single}
\fmfleft{in,p1}
\fmfright{out,p2}
\fmfdot{c}
\fmf{dashes_arrow,label=\small{new}}{in,c}
\fmf{dashes_arrow}{c,out}
\fmf{plain_arrow,right, tension=0.8, label=\small{lables}}{c,c}
\fmf{phantom, tension=0.2}{p1,p2}
\end{fmfgraph*}


\begin{fmfgraph*}(80,40)
\fmfkeep{schannel}
\fmfleft{i1,i2}
\fmfright{o1,o2}
\fmf{plain}{i1,v1}
\fmf{plain}{i2,v1}
\fmf{plain,left=0.5,tension=0.4}{v1,v2}
\fmf{plain,right=0.5,tension=0.4}{v1,v2}
\fmf{plain}{v2,o1}
\fmf{plain}{v2,o2}
\fmfdot{v1,v2}
\end{fmfgraph*}

\begin{fmfgraph*}(80,40)
\fmfkeep{tchannel}
\fmfleft{i1,i2}
\fmfright{o1,o2}
\fmf{plain}{i1,v1,o1}
\fmf{plain}{i2,v2,o2}
\fmf{plain,left=0.5,tension=0.4}{v1,v2}
\fmf{plain,right=0.5,tension=0.4}{v1,v2}
\fmfdot{v1,v2}
\end{fmfgraph*}

\begin{fmfgraph*}(80,40)
\fmfkeep{uchannel}
\fmfleft{i1,i2}
\fmfright{o1,o2}
\fmf{plain}{i1,v1}
\fmf{phantom}{v1,o1} 
\fmf{plain}{i2,v2}
\fmf{phantom}{v2,o2} 
\fmf{plain,left=0.5,tension=0.4}{v1,v2}
\fmf{plain,right=0.5,tension=0.4}{v1,v2}
\fmf{plain,tension=0}{v1,o2}
\fmf{plain,tension=0}{v2,o1}
\fmfdot{v1,v2}
\end{fmfgraph*}

\begin{fmfgraph*}(80,40)
\fmfkeep{tadpolesix}
\fmfbottom{i1,o1}
\fmftop{i2,o2}
\fmf{plain}{i1,v1,o1}
\fmf{plain}{i2,v1,o2}
\fmf{plain,right=90,tension=0.8}{v1,v1}
\fmfdot{v1}
\end{fmfgraph*}


\begin{fmfgraph*}(100,36)
\fmfkeep{bubble}
\fmfleft{in}
\fmfright{out}
\fmfdot{v1}
\fmfdot{v2}
\fmf{plain}{in,v1}
\fmf{plain}{v2,out}
\fmf{plain,left,tension=0.6}{v1,v2}
\fmf{plain,right,tension=0.6}{v1,v2}
\end{fmfgraph*}


\begin{fmfgraph*}(100,36)
\fmfkeep{tadpole}
\fmfset{dash_len}{6pt} 
\fmfleft{in,p1}
\fmfright{out,p2}
\fmfdot{c}
\fmf{plain}{in,c}
\fmf{plain}{c,out}
\fmf{plain,right, tension=0.8}{c,c}
\fmf{phantom, tension=0.2}{p1,p2}
\end{fmfgraph*}

\end{fmffile}

}

\hfill{Imperial-TP-LW-2014-03}

\vspace{40pt}

\begin{center}
{\huge{\bf One- and two-loop checks for the $\mathbf{AdS_3\times S^3\times T^4}$ superstring with mixed flux 
\vskip 6pt
 }}

\vspace{20pt}

{\bf \large  Per Sundin$^a$} and {\bf\large Linus Wulff$\,^b$}

\vspace{15pt}

{$^a$ \it\small Universit\'a di Milano-Bicocca and INFN Sezione di Milano-Bicocca,\\\hspace{.5cm} Dipartimento de Fisica,
Piazza della Scienza 3, I-20126 Milano, Italy}\\
{$^b$ \it\small The Blackett Laboratory, Imperial College, London SW7 2AZ, U.K.}\\

\vspace{50pt}

{\bf Abstract}

\end{center}
\noindent
We compute the one-loop worldsheet S-matrix for the superstring in $AdS_3\times S^3\times T^4$ supported by a combination of RR and NSNS flux in the massive sector. In the appropriate regularization scheme it agrees with the S-matrix found from symmetry considerations including the proposed dressing phases. As previously observed in other cases the massless modes decouple from the calculation of one-loop massive mode scattering. We also calculate the correction to the dispersion relation at two loops in the Near-Flat-Space limit and find agreement with a proposal for the form of the exact dispersion relation for the massive modes. Somewhat surprisingly the massless modes again decouple from the calculation (in a suitable regularization scheme). We also compute the correction to the dispersion relation for the massless modes and compare to a suggestion for the exact dispersion relation, finding a small discrepancy. The corresponding calculations for $AdS_5\times S^5$ and $AdS_2\times S^2\times T^6$ are used for comparison.

\vskip 90pt
\pagebreak
\setcounter{page}{1}


\section{Introduction and summary}\label{sec:intro}
Several instances of the $AdS/CFT$-correspondence have an underlying integrable structure. Here we will focus on the AdS-side of the correspondence. In the classic example we have strings in $AdS_5\times S^5$ \cite{Metsaev:1998it} whose (classical) integrability was demonstrated in \cite{Bena:2003wd}. By now many more integrable examples are known. In order of decreasing number of supersymmetries we have; $AdS_4\times\mathbbm{CP}^3$ \cite{Arutyunov:2008if,Stefanski:2008ik,Gomis:2008jt,Sorokin:2010wn} preserving 24 supersymmetries, $AdS_3\times S^3\times T^4$ and $AdS_3\times S^3\times S^3\times S^1$ \cite{Babichenko:2009dk,Sundin:2012gc,Cagnazzo:2012se} preserving 16 supersymmetries, $AdS_2\times S^2\times T^6$ \cite{Sorokin:2011rr} and $AdS_2\times S^2\times S^2\times T^4$, $AdS_3\times S^2\times S^2\times T^3$ etc. \cite{Wulff:2014kja} preserving 8 supersymmetries.\footnote{Reviews of the integrability machinery applied to some of these backgrounds include \cite{Arutyunov:2009ga,Beisert:2010jr} for $AdS_5\times S^5$, \cite{Klose:2010ki} for $AdS_4\times\mathbbm{CP}^3$ and \cite{Sfondrini:2014via} for $AdS_3\times S^3\times M_4$.}  Even more examples can be obtained by dualities or deformations of these. With such a rich set of examples it is natural to try to find the structures they all have in common. In this paper we will focus on the examples of the form $AdS_n\times S^n\times T^{10-2n}$ where $n=2,3,5$. The reason for this is that they are the simplest examples for semi-classical computations in a near-BMN expansion since they do not have any cubic interaction terms in the Lagrangian. 

In \cite{Roiban:2014cia} the one-loop worldsheet S-matrix in the massive sector was computed for these examples and the dressing phases were shown to match the conjectured ones. In that paper the backgrounds were supported by pure RR flux. Here we will complete the one-loop analysis by considering also $AdS_3\times S^3\times T^4$ with a mix of RR and NSNS flux. We find that this case, although somewhat more complicated, exhibits the same wave function renormalization as the pure RR backgrounds of \cite{Roiban:2014cia}. We also show that just as for the pure RR cases one can find a regularization scheme such that the massless modes decouple and the full one-loop result follows from the supercoset sigma model. The one-loop dressing phases again agree with the ones found via generalized unitarity and crossing symmetry \cite{Engelund:2013fja,Babichenko:2014yaa,Bianchi:2014rfa}.

We also compute the two-loop correction to the dispersion relation for strings in $AdS_3\times S^3\times T^4$ with mixed flux and find that, for the massive modes, it agrees with the proposed exact form \cite{Hoare:2013lja,Lloyd:2014bsa}. The computation is done in the near-flat-space limit to minimize the complexity of the calculations. Surprisingly we find again that there is a regularization scheme in which the massless modes decouple from the calculation and the supercoset model gives the complete answer. The same is shown to happen in $AdS_2\times S^2\times T^6$. For the correction to the dispersion relation for massless modes however, we find a result which differs from the suggested form for the exact dispersion relation \cite{Borsato:2014exa,Borsato:2014hja,Lloyd:2014bsa}. This discrepancy is not limited to the mixed flux case but is there already in the pure RR case (it takes the same form in $AdS_2\times S^2\times T^6$ but in that case the exact dispersion relation is not known). We give some possible explanations for this mismatch below.

The results for the two-loop correction to the dispersion relation in the Near-Flat-Space (NFS) limit can be summarized as follows
\begin{align}
\label{eq:nfs-dispersion-relation}
AdS_5\times S^5:&\qquad\varepsilon^2|_{2-loop}=-\frac{p_-^4}{192}\,,\nonumber\\
AdS_3\times S^3\times T^4:&\qquad\varepsilon^2|_{2-loop}=
\left\{\begin{array}{cc}
-\hat q^2\frac{p_-^4}{192} & (\mbox{massive})\\
-\hat q^2\frac{p_-^4}{32\pi^2} & (\mbox{massless})
\end{array}
\right.\,,\\
AdS_2\times S^2\times T^6:&\qquad\varepsilon^2|_{2-loop}=
\left\{
\begin{array}{cc}
-\frac14\frac{p_-^4}{192} & (\mbox{massive})\\
-\frac12\frac{p_-^4}{32\pi^2} & (\mbox{massless})
\end{array}
\right.\,.\nonumber
\end{align}
The calculation for $AdS_5\times S^5$ was first performed in \cite{Klose:2007rz} and agrees with the expansion of the exact dispersion relation
\begin{equation}
\varepsilon^2=1+4h^2\sin^2\frac{\mathrm p}{2}\,.
\label{eq:disp-ads5}
\end{equation} 
This case is included here only to demonstrate the similarities to the other two cases. For the $AdS_3\times S^3\times T^4$ case the correction to the dispersion relation for the massive modes agrees with the suggested exact dispersion relation \cite{Hoare:2013lja,Lloyd:2014bsa}
\begin{equation}
\varepsilon_\pm^2=(qg\mathrm p\pm\mathrm m)^2+4\hat q^2h^2\sin^2\frac{\mathrm p}{2}\,,
\label{eq:disp-ads3}
\end{equation} 
with $\mathrm m=1$. Here $g$ is the string tension (this was replaced by $h$ in \cite{Hoare:2013lja}) and the parameter $0\leq q\leq1$ controls the amount of NSNS-flux\footnote{The exact dispersion relation for the case of pure RR-flux, i.e. $q=0$, was analyzed in \cite{Borsato:2014exa,Borsato:2014hja}.} while $\hat q^2=1-q^2$. Surprisingly we find that, in an appropriate regularization scheme, the massless modes completely decouple from this calculation and the result is simply what one obtains from the supercoset sigma-model. The fact that massless modes decouple, in a certain regularization scheme, from computations of two and four-point functions for massive modes was seen at one loop in \cite{Roiban:2014cia}. Our results suggest that this decoupling might extend to two loops. This is somewhat surprising since the supercoset sigma-model is a non-critical string and one would expect problems unless the massless modes are included to make the model into the full critical superstring. These problems could of course show up only at three loops.

For the massless modes the exact dispersion relation was suggested to have the same form as in (\ref{eq:disp-ads3}) but with $\mathrm m=0$ \cite{Lloyd:2014bsa}. But here we find a result that differs by a factor of $6/\pi^2$ from the one coming from the suggested exact dispersion relation. The extra factor of $1/\pi^2$ can be traced to the types of integrals that contribute in this case, which have one massless and two massive modes running in the loops. These are very different to the integrals with three massive modes (of the same mass) running in the loops that appear in the calculation for the massive modes and this is the origin of the relative factor of $1/\pi^2$. This discrepancy does not appear to be an artifact of the NFS limit since the types of integrals that appear are essentially the same in the full BMN case. It is also difficult to imagine that one would get a different answer by first performing the full BMN calculation and then taking the NFS limit of the result. The only possible caveat here is that there will be regularization issues in the BMN case, related to UV and/or IR divergences, which are not present in the NFS case, but the contribution from these will come with lower powers of $p_-$ and it is therefore difficult to see how that could change the result. Furthermore, this does not seem to be a likely cause of the mismatch since the full two-loop two-point function is UV-finite, as we demonstrate in appendix \ref{sec:massless-bmn}. Finally, it is interesting to note the factor of $6$. In the massive case a factor of $1/6$ comes from expanding the sine in the exact dispersion relation. The extra factor of $6$ in the massless case might therefore suggest a dispersion relation for these modes which is not of the $\sin^2$-form.

Let us briefly remark on the derivation of the exact dispersion relation. One starts from the usual light-cone gauge BMN string and relaxes the level-matching condition. This leads to an off-shell (in the sense of the level-matching condition) symmetry algebra which is that of the BMN vacuum with an additional central extension \cite{Arutyunov:2006ak}. In the $AdS_3\times S^3\times T^4$ case this algebra is centrally extended $\mathfrak{psu}(1|1)^4$. The massive modes arrange themselves into two short multiplets of this symmetry algebra consisting of two bosons and two fermions each. These have mass parameter $\mathrm m=\pm1$ respectively. The shortening condition (we take $q=0$ for simplicity)
\begin{equation}
H^2=M^2+4C\bar C\,,
\label{eq:shortcond}
\end{equation}
where $H$ is the Hamiltonian and $M,C$ central charges, gives the dispersion relation provided the central charges take the form 
\begin{equation}
M=\mathrm m\qquad\mbox{and}\qquad C=\frac{ih\zeta}{2}(e^{iP}-1)\,,
\label{eq:centralcharges}
\end{equation}
with $\zeta$ a phase and $P$ the worldsheet momentum. This form for the central charges (with $h=g$, the string tension) follows from a classical calculation of the Poisson bracket of two supercharges.

The situation for the massless modes is similar, but with some differences. There are four massless bosons and four massless fermions and they can again be arranged into two short multiplets of the off-shell symmetry algebra \cite{Borsato:2014exa,Borsato:2014hja,Lloyd:2014bsa}. In this case however the two multiplets have the same value of the parameter $\mathrm m$ labeling the representation, namely $\mathrm m=0$. The off-shell symmetry algebra also has an $\mathfrak{su}(2)$-factor\footnote{
This extra symmetry is manifest only in the type IIB case, which is the case usually considered. Here we work in type IIA since the $AdS_2\times S^2\times T^6$ solution we consider takes a nicer form in type IIA and this allows us to treat the $AdS_3$ and $AdS_2$ case on the same footing. This is however not an issue when comparing to the literature since we consider only the massive S-matrix at one loop and at two loops we work in the Near-Flat-Space limit in which the type IIA and type IIB string actions become identical, see sec. \ref{sec:NFS}.} which commutes with $\mathfrak{psu}(1|1)^4$ and the massless modes form a doublet of the $\mathfrak{su}(2)$. Their transformation under this $\mathfrak{su}(2)$ is the reason that the massless modes must sit in a short multiplet which is an $\mathfrak{su}(2)$ doublet, rather than a long multiplet which is an $\mathfrak{su}(2)$ singlet. Since the shortening condition is the same as before (\ref{eq:shortcond}) one obtains the dispersion relation of \cite{Borsato:2014exa,Borsato:2014hja} provided the central charges take the same form (\ref{eq:centralcharges}). This form of the central charges was verified at the classical level in \cite{Borsato:2014hja,Lloyd:2014bsa}.

Therefore, in order to reconcile our findings with those of \cite{Borsato:2014hja,Lloyd:2014bsa} we can speculate on the following two possible reasons for the discrepancy:
\begin{itemize}
	\item[(i)] The form of the central charges (\ref{eq:centralcharges}) receives quantum corrections (beyond corrections to $h$) in the case of massless modes.
	\item[(ii)] The asymptotic states for the massless modes differ by a momentum-dependent factor between the two computations. 
\end{itemize}
To check whether it's the second of these one should calculate a two-loop S-matrix element. This is beyond the scope of the present paper. Note that we do not expect this discrepancy to be an artifact of regularization. The NFS calculation is finite and in fact we demonstrate in appendix \ref{sec:massless-bmn} that the same is true for the full BMN computation.

In $AdS_2\times S^2\times T^6$ the situation is less understood. There is no known way to derive the exact dispersion relation in this case, even for the massive modes, since they are in long representations of the BMN vacuum symmetry algebra $\mathfrak{psu}(1|1)^2\ltimes\mathbbm R^3$ \cite{Hoare:2014kma}. This is in contrast to the $AdS_5\times S^5$ and $AdS_3\times S^3\times T^4$ cases where the massive modes are in short representations and the shortening condition gives rise to the exact dispersion relation. However, the calculation of the two-loop correction to the dispersion relation for the massive modes, which was first performed in \cite{Murugan:2012mf}, gives a result very similar to the other cases suggesting that the exact dispersion relation should take a similar form also in this case. Notice however the extra factor of $\frac14$ compared to the other cases (this factor was hidden in \cite{Murugan:2012mf} due to an unfortunate choice of $g/2$ instead of $g$ for the string tension). A similar difference was noticed at one loop in \cite{Roiban:2014cia} suggesting that, roughly, the effective coupling in this case is $2g$ rather than $g$. This should also be related to the factors of 2 appearing in the dressing phase \cite{Abbott:2013kka}.

For the massless modes the result is the same as in the $AdS_3$ case (with $q=0$) except for a factor of $\frac12$. Therefore the idea of the effective coupling being $2g$ rather than $g$ appears to only be true in the massive sector. Just as in $AdS_3$ the massive and massless correction differ by factors of $1/\pi^2$, the reason again being that one contribution comes from integrals with only massive modes in the loop while the other comes from integrals with one massless and two massive modes in the loop.

The outline of the paper is as follows. In section \ref{sec:action} we describe the superstring action and the form of the $AdS_3\times S^3\times T^4$ background with mixed fluxes. We then consider the near BMN expansion and the NFS limit in section \ref{sec:BMN}. Section \ref{sec:reg} outlines the regularization scheme we employ for the loop computations, while sections \ref{sec:twopoint} and \ref{sec:Smatrix} give the results for the one and two loop two-point functions and the one-loop S-matrix respectively. We end with some conclusions. Appendix \ref{app:NFSids} describes the integral identities used in the two-loop computation and appendix \ref{sec:massless-bmn} demonstrates the finiteness of the full BMN two-loop two-point function for the massless modes.

\section{Superstring action}\label{sec:action}
The Green-Schwarz superstring action in any supergravity background can be expanded  in powers of  fermions as (here $g$ denotes the string tension)
\begin{equation}
S=g\int d^2\xi\,(\mathcal L^{(0)}+\mathcal L^{(2)}+\ldots)\,,
\end{equation}
and in a general type II supergravity background the action is known up to quartic order \cite{Wulff:2013kga}. The purely bosonic terms in the Lagrangian are 
\begin{equation}
\label{eq:L0}
\mathcal L^{(0)}=\frac12\gamma^{ij}e_i{}^ae_j{}^b\eta_{ab}+\frac12\varepsilon^{ij}B^{(0)}_{ij}\,,\qquad\gamma^{ij}=\sqrt{-h}h^{ij}\,,
\end{equation}
where $e_i{}^a$ $(a=0,\ldots,9;\,i=0,1)$ denotes the bosonic vielbein pulled back to the worldsheet and $B_{ij}^{(0)}=e_i{}^ae_j{}^bB^{(0)}_{ab}$ is (the pull-back of) the  lowest component in the Grassmann parameter $\Theta$-expansion of the NSNS two-form superfield $B$. For the terms involving fermions we will write the expressions appropriate to type IIA supergravity, taking $\Theta^\alpha$ to be a 32-component Majorana spinor.
 
The terms quadratic in fermions take the form
\begin{equation}
\mathcal L^{(2)}=\frac{i}{2}e_i{}^a\,\Theta\Gamma_aK^{ij}\mathcal D_j\Theta\,,\qquad K^{ij}=\gamma^{ij}-\varepsilon^{ij}\Gamma_{11}\,,
\end{equation}
where
\begin{equation}
\label{eq:DbA}
\mathcal D\Theta=
\big(d-\frac{1}{4}\omega^{ab}\Gamma_{ab}+\frac{1}{8}e^aM_a\big)\Theta\,,\qquad M_a=H_{abc}\,\Gamma^{bc}\Gamma_{11}+\mathcal S\Gamma_a\,.
\end{equation}
Here  $\omega^{ab}$ is the spin connection, $H=dB$ is the NSNS three-form field strength
and the RR fields enter the action through the bispinor\footnote{Here $\phi$ is the dilaton and we use the convention $F^{(n)} = \frac{1}{n!}dx^{m_n}\wedge\dots\wedge dx^{m_1}F_{m_1\dots m_n}$ for the form fields.}
\begin{equation}
\label{eq:SbA}
\mathcal S=e^\phi\big(\frac12F^{(2)}_{ab}\Gamma^{ab}\Gamma_{11}+\frac{1}{4!}F^{(4)}_{abcd}\Gamma^{abcd}\big)\,.
\end{equation}

The quartic fermion terms take the form \cite{Wulff:2013kga}
\begin{align}
\mathcal L^{(4)}=&
-\frac{1}{8}\Theta\Gamma^a\mathcal D_i\Theta\,\Theta\Gamma_aK^{ij}\mathcal D_j\Theta
+\frac{i}{24}e_i{}^a\,\Theta\Gamma_aK^{ij}\mathcal M\mathcal D_j\Theta
+\frac{i}{192}e_i{}^ae_j{}^b\,\Theta\Gamma_aK^{ij}(M+\tilde M)\mathcal S\Gamma_b\Theta
\nonumber\\
&{}
+\frac{1}{192}e_i{}^ce_j{}^d\,\Theta\Gamma_c{}^{ab}K^{ij}\Theta\,(3\Theta\Gamma_dU_{ab}\Theta-2\Theta\Gamma_aU_{bd}\Theta)
\nonumber\\
&{}
-\frac{1}{192}e_i{}^ce_j{}^d\,\Theta\Gamma_c{}^{ab}\Gamma_{11}K^{ij}\Theta\,(3\Theta\Gamma_d\Gamma_{11}U_{ab}\Theta+2\Theta\Gamma_a\Gamma_{11}U_{bd}\Theta)\,,
\label{eq:L4}
\end{align}
where the new objects appearing in this expression are defined as
\begin{align}
\mathcal M^\alpha{}_\beta=&
M^\alpha{}_\beta+\tilde M^\alpha{}_\beta
+\frac{i}{8}(M^a\Theta)^\alpha\,(\Theta\Gamma_a)_\beta
-\frac{i}{32}(\Gamma^{ab}\Theta)^\alpha\,(\Theta\Gamma_aM_b)_\beta
-\frac{i}{32}(\Gamma^{ab}\Theta)^\alpha\,(C\Gamma_aM_b\Theta)_\beta, 
\nonumber\\
M^\alpha{}_\beta=&
\frac12\Theta T\Theta\,\delta^\alpha_\beta
-\frac12\Theta\Gamma_{11}T\Theta\,(\Gamma_{11})^\alpha{}_\beta
+\Theta^\alpha\, (CT\Theta)_\beta
+(\Gamma^aT\Theta)^\alpha\,(\Theta\Gamma_a)_\beta, 
\label{eq:T}\\
T=&\frac{i}{2}\nabla_a\phi\,\Gamma^a+\frac{i}{24}H_{abc}\,\Gamma^{abc}\Gamma_{11}+\frac{i}{16}\Gamma_a\mathcal S\Gamma^a\,,\quad
U_{ab}=\frac{1}{4}\nabla_{[a}M_{b]}+\frac{1}{32}M_{[a}M_{b]}-\frac{1}{4}R_{ab}{}^{cd}\,\Gamma_{cd}\,,\nonumber
\end{align}
and $\tilde M=\Gamma_{11}M\Gamma_{11}$. For a supergravity background preserving some supersymmetries the dilatino equation is $T\xi=0$ and the integrability condition for the gravitino equation is $U_{ab}\xi=0$, where $\xi$ is a Killing spinor \cite{Wulff:2013kga,Wulff:2014kja}.

The background we are interested in here is (type IIA) $AdS_3\times S^3\times T^4$ supported by a mix of RR and NSNS flux (we use the conventions of \cite{Wulff:2014kja})
\begin{equation}
\label{eq:fluxes}
H=2q(\Omega_{AdS_3}+\Omega_{S^3})\,,\qquad F^{(4)}=2\hat qe^{-\phi}dx^9(\Omega_{AdS_3}+\Omega_{S^3})\,,\qquad\mbox{with}\qquad q^2+\hat q^2=1\,.
\end{equation}
Here the $AdS_3$ and $S^3$ radii have been set to 1 and the dilaton $\phi$ is constant. From (\ref{eq:SbA}) and (\ref{eq:T}) we find 
\begin{equation}
\mathcal S=-4\hat q\mathcal P_{16}\Gamma^{0129}\,,\qquad T=-\frac{i}{2}\Gamma^{012}[\hat q\Gamma^9+2q\Gamma_{11}](1-\mathcal P_{16})\,,\qquad\mathcal P_{16}=\frac12(1+\Gamma^{012345})\,.
\end{equation}
The projection operator $\mathcal P_{16}$ projects onto the 16 supersymmetries preserved by the background. Accordingly we can split the fermions into $16+16$ as $\vartheta=\mathcal P_{16}\Theta$ and $\upsilon=(1-\mathcal P_{16})\Theta$ which we will refer to as coset and non-coset fermions respectively since only the former appear in the supercoset sigma model. These are also the ones that transform under the supersymmetries.

For the background space-time we take the metric of $AdS_3$ to be
\begin{equation}
ds^2_{AdS_3}=-\Big(\frac{1+\frac12|z|^2}{1-\frac12|z|^2}\Big)^2dt^2+\frac{2|dz|^2}{(1-\frac12|z|^2)^2}\,,
\label{eq:AdSmetric}
\end{equation}
with the spatial coordinates grouped together into one complex coordinate. Similarly,  the $S^3$ metric is
\begin{equation}
ds^2_{S^3}=\Big(\frac{1-\frac12|y|^2}{1+\frac12|y|^2}\Big)^2d\varphi^2+\frac{2|dy|^2}{(1+\frac12|y|^2)^2}\,.
\label{eq:Smetric}
\end{equation}
Since the NSNS two-form potential $B$ appears explicitly in the action we will need its form in these coordinates. From (\ref{eq:fluxes}) we see that its three-form field strength takes the form
\begin{equation}
H=dB=-2iq\frac{1+\frac12|z|^2}{(1-\frac12|z|^2)^3}dzd\bar zdt+2iq\frac{1-\frac12|y|^2}{(1+\frac12|y|^2)^3}dyd\bar yd\varphi\,.
%
\end{equation}
It is not hard to show that we can take
\begin{equation}
B=-iq\frac{zd\bar z-\bar zdz}{(1-\frac12|z|^2)^2}dt+iq\frac{yd\bar y-\bar ydy}{(1+\frac12|y|^2)^2}d\varphi\,.
\end{equation}
The action is rather involved and in order to handle it we will consider a perturbative expansion around the classical solution given by the BMN geodesic.

\section{Near--BMN expansion}\label{sec:BMN}
We wish to expand the superstring action described in the previous section around the BMN solution $t=\varphi=\tau$ \cite{Berenstein:2002jq}. At the same time we will fix the light-cone gauge and the corresponding kappa symmetry gauge
\begin{equation}
x^+=\tau\,,\qquad\Gamma^+\Theta=0\,,
\end{equation}
where $x^\pm=\frac12(t\pm\varphi)$. The complete gauge fixing includes also the conditions
\begin{equation}
p^+\equiv-\frac12\frac{\partial\mathcal L}{\partial\dot x^-}=1\,,\qquad\frac{\partial\mathcal L}{\partial {x^-}'}=0\,,
\end{equation}
which can be used to solve for the higher corrections to the worldsheet metric $\gamma^{ij}=\eta^{ij}+\hat\gamma^{ij}$. The Virasoro constraints can be used to solve for $x^-$ in terms of the other fields, but its form will not be needed here.

In this gauge the Lagrangian has an expansion in the number of transverse fields
\begin{equation}
\mathcal L=\mathcal L_2+\frac1g\mathcal L_4+\frac{1}{g^2}\mathcal L_6+\ldots\,.
\end{equation}
The absence of terms with an odd number of fields is a feature of $AdS_n\times S^n$ backgrounds which simplifies the analysis at loop level as it leads to fewer Feynman diagrams. The quadratic Lagrangian takes the form
\begin{align}
\label{eq:L2}
\mathcal{L}_2=&
|\partial_i\tilde z|^2-\hat q^2|\tilde z|^2
+|\partial_i\tilde y|^2-\hat q^2|\tilde y|^2
+|\partial_i u_1|^2+|\partial_i u_2|^2
+i\bar\chi_L^r\partial_-\chi_L^r
+i\bar\chi_R^r\partial_+\chi_R^r
-\hat q\bar\chi_L^r\chi_R^r
\nn\\
&{}
-\hat q\bar\chi_R^r\chi_L^r
+i\bar\chi_L^{r'}\partial_-\chi_L^{r'}
+i\bar\chi_R^{r'}\partial_+\chi_R^{r'}\,,
\end{align}
where $r=1,2$ runs over the massive fermions, $r'=3,4$ over the massless ones and $u_{1,2}$ denotes the two complex massless bosons coming from $T^4$. The spectrum can be summarized as follows:
\begin{align}
m=\hat q:&\qquad\qquad\mbox{Bosons: }\tilde z,\,\tilde y &\mbox{Fermions: }\chi^1,\,\chi^2\phantom{\,.}
\nn\\
m=0:&\qquad\qquad\mbox{Bosons: }u_1,\,u_2 &\mbox{Fermions: }\chi^3,\,\chi^4\,.
\end{align}
To put the quadratic Lagrangian in the Lorentz-invariant form (\ref{eq:L2}) we have performed the following, $\sigma$-dependent, $U(1)$-rotation
\begin{equation}
\label{eq:quadratic-shift}
z=e^{-iq\sigma}\tilde z\,,\quad y=e^{iq\sigma}\tilde y
\end{equation}
of the massive bosons. This transformation is accompanied by the transformation 
\begin{equation}
\vartheta=e^{-q\sigma\Gamma_{12}}\tilde\vartheta
\end{equation}
of the coset fermions. The  coset and non-coset fermions, satisfying $\vartheta=\frac12(1+\Gamma^{1234})\Theta$ and $\upsilon=\frac12(1-\Gamma^{1234})\Theta$ after the kappa symmetry gauge fixing,
give rise to massive and massless modes respectively, i.e.
\begin{equation}
\tilde\vartheta\rightarrow\chi^1,\,\chi^2\qquad\upsilon\rightarrow\chi^3\,,\chi^4\,.
\end{equation}
Note that we could of course have chosen to work with the original fields, $y,\,z$ etc., but the non-Lorentz-invariant form of the quadratic action would then greatly complicate loop calculations.

The interaction terms are quite complicated and we will only give the quartic bosonic terms which take the form
\begin{eqnarray}
\mathcal L_4^B&=&
\frac{\hat q^2}{2}|\tilde z|^2\big(|\dot{\tilde z}|^2-|\tilde z'|^2\big)
-\frac{\hat q^2}{2}|\tilde y|^2\big(|\dot{\tilde y}|^2-|\tilde y'|^2\big)
-\frac{q^2\hat q^2}{2}\big(|\tilde z|^4-|\tilde y|^4\big)
\nonumber\\
&&{}
-\frac12\big(\hat q^2|\tilde z|^2-\hat q^2|\tilde y|^2-iq\tilde z'\bar{\tilde z}-iq\tilde y'\bar{\tilde y}\big)
\big(
|\dot{\tilde z}|^2+|\tilde z'|^2
+|\dot{\tilde y}|^2+|\tilde y'|^2
+|\dot u_I|^2+|u'_I|^2
\big)
\nonumber\\
&&{}
-\frac{iq}{2}\big(\dot{\tilde z}\bar{\tilde z}+\dot{\tilde y}\bar{\tilde y}\big)
\big(
\dot{\tilde z}\bar{\tilde z}'+\tilde z'\dot{\bar{\tilde z}}
+\dot{\tilde y}{\bar{\tilde y}}'+\tilde y'\dot{\bar{\tilde y}}
+\dot u_I\bar u'_I+u'_I\dot{\bar u}_I
\big)
\nonumber\\
&&{}
-iq\hat q^2\big(|\tilde z|^2-|\tilde y|^2)(\tilde z'\bar{\tilde z}-\tilde y'\bar{\tilde y}\big)
-\frac{iq\hat q^2}{2}\big(|\tilde z|^2+|\tilde y|^2)(\tilde z'\bar{\tilde z}+\tilde y'\bar{\tilde y}\big)
\nonumber\\
&&{}
+\frac{q^2}{2}\big[(\dot{\tilde z}\bar{\tilde z})^2-(\tilde z'\bar{\tilde z})^2-(\dot{\tilde y}\bar{\tilde y})^2+(\tilde y'\bar{\tilde y})^2\big]
+\mbox{c.c.}
\end{eqnarray}

\subsection{Near-Flat-Space limit}\label{sec:NFS}
Our one-loop calculations will be done with the full BMN action but at two loops this becomes highly involved. In order to simplify the model so that we can carry out computations at two loops as well we will take the so-called Near-Flat-Space (NFS) limit \cite{Maldacena:2006rv}. It consists of performing the following chiral rescalings, or boosts, in the (non-Lorentz-invariant) BMN Lagrangian 
\begin{equation}
\partial_\pm\rightarrow g^{\mp1/2}\partial_\pm\,,\qquad\tilde\Theta_\pm\rightarrow g^{\mp1/4}\tilde\Theta_\pm\,,
\end{equation}
where $\Theta_\pm=\frac12(1\pm\Gamma_{11})\Theta$. This leaves Lorentz-invariant terms invariant but rescales non-Lorentz-invariant terms. Keeping only the leading $g^0$-terms gives
\begin{eqnarray}
\label{eq:LNFS}
\mathcal L_{NFS}=\mathcal L_2+\hat q^2\mathcal L_4^{(NFS)}\,,
\end{eqnarray}
where $\mathcal L_2$ is the same as before (\ref{eq:L2}). Note that if we had not first put the quadratic action in a Lorentz-invariant form by introducing the tilded variables we would obtain a very different NFS limit. The quartic terms take the form
\begin{align}
\mathcal L_4^{(NFS)}=&
-\frac14\big(|\tilde z|^2-|\tilde y|^2\big)\big(|\partial_-\tilde z|^2+|\partial_-\tilde y|^2+|\partial_-u_I|^2
-i\tilde\vartheta_-\Gamma^-\partial_-\tilde\vartheta_-
-i\upsilon_-\Gamma^-\partial_-\upsilon_-
\big)
\nonumber\\
&{}
+\frac14\big(\partial_-\tilde z\bar{\tilde z}+\partial_-\tilde y\bar{\tilde y}\big)\tilde\vartheta_-\Gamma^{12-}\tilde\vartheta_-
+\frac14\big(\partial_-\tilde z\bar{\tilde z}-\partial_-\tilde y\bar{\tilde y}\big)\upsilon_-\Gamma^{12-}\upsilon_-
\nonumber\\
&{}
-\frac{i}{4}\partial_-(\tilde z\bar{\tilde y})\tilde\vartheta_-(\Gamma^1+i\Gamma^2)\Gamma^{3-}\tilde\vartheta_-
-\frac{i}{4}\partial_-(\tilde z\tilde y)\upsilon_-(\Gamma^1+i\Gamma^2)\Gamma^{3-}\upsilon_-
\nonumber\\
&{}
+\frac{i}{4}\tilde z\,\upsilon_-[\partial_-u_1(\Gamma^6+i\Gamma^7)+\partial_-u_2(\Gamma^8+i\Gamma^9)+\mbox{c.c.}](\Gamma^1+i\Gamma^2)\Gamma^-\tilde\vartheta_-
\nonumber\\
&{}
-\frac{i}{4}\tilde y\,\upsilon_-[\partial_-u_1(\Gamma^6+i\Gamma^7)+\partial_-u_2(\Gamma^8+i\Gamma^9)+\mbox{c.c.}](\Gamma^3+i\Gamma^4)\Gamma^-\tilde\vartheta_-
+\mbox{c.c.}
\nonumber\\
&{}
+\frac{1}{6}\tilde\vartheta_-\Gamma^{12-}\tilde\vartheta_-\,\upsilon_-\Gamma^{12-}\upsilon_-
+\frac{1}{12}\upsilon_-\Gamma^{a'1-}\tilde\vartheta_-\,\upsilon_-\Gamma^{a'1-}\tilde\vartheta_-
+\frac{1}{12}\upsilon_-\Gamma^{a'2-}\tilde\vartheta_-\,\upsilon_-\Gamma^{a'2-}\tilde\vartheta_-
\nonumber\\
&{}
-\frac{1}{12}\upsilon_-\Gamma^{a'3-}\tilde\vartheta_-\,\upsilon_-\Gamma^{a'3-}\tilde\vartheta_-
-\frac{1}{12}\upsilon_-\Gamma^{a'4-}\tilde\vartheta_-\,\upsilon_-\Gamma^{a'4-}\tilde\vartheta_-\,,
\label{eq:L4NFS}
\end{align}
where $a'=6,7,8,9$ runs over the $T^4$-directions. We have kept the fermions in 10d notation to shorten the expression. Recall that the massive fermions come from $\tilde\vartheta$ and the massless ones from $\upsilon$.

It is interesting to note that $q$, the parameter controlling the amount of NSNS flux, enters only in the masses in (\ref{eq:L2}) and in the quartic coupling in (\ref{eq:LNFS}) as $\hat q^2=1-q^2$. Therefore the calculation for $q\neq0$ is almost identical to the $q=0$ calculation. The exception is $q=1$, i.e. $AdS_3\times S^3\times T^4$ with pure NSNS flux, in which case the NFS Lagrangian becomes quadratic and all modes become massless.

Note that so far we have been talking about the type IIA $AdS_3\times S^3\times T^4$ supergravity solution. In this case the (local) $SO(4)$-symmetry of the $T^4$ is broken to $U(1)$ by the fluxes (\ref{eq:fluxes}). However, we can easily go to the more symmetric type IIB case by T-dualizing along the 9-direction. The T-duality transformation is
\begin{equation}
\partial_+u_2\rightarrow\partial_+u_2\,,\quad
\partial_-u_2\rightarrow\partial_-\bar u_2\,,\qquad
\Theta_+\rightarrow\Theta^1\,,\quad
\Theta_-\rightarrow\Gamma^9\Theta^2\,.
\end{equation}
This transformation is easily seen to leave the NFS Lagrangian (\ref{eq:LNFS}) and (\ref{eq:L4NFS}) unchanged, which means that the NFS Lagrangian is the same in the type IIA and type IIB case.

The $SO(4)$-symmetry coming from the $T^4$ splits into $SU(2)_L\times SU(2)_R$ which act as follows. We have two $SU(2)_R$ doublets
\begin{equation}
\left(\begin{array}{c}
u_1\\
u_2
\end{array}\right)
\qquad\mbox{and}\qquad
\left(\begin{array}{c}
\Gamma^6\upsilon_-^-\\
\Gamma^8\upsilon_-^-
\end{array}\right)
\end{equation}
and two $su(2)_L$ doublets
\begin{equation}
\left(\begin{array}{c}
u_1\\
\bar u_2
\end{array}\right)
\qquad\mbox{and}\qquad
\left(\begin{array}{c}
\Gamma^6\tilde\vartheta_-^-\\
\Gamma^8\tilde\vartheta_-^-
\end{array}\right)\,,
\end{equation}
where $\tilde\Theta_-^\pm=\frac12(1\pm i\Gamma^{67})\tilde\Theta_-$. Note that the massless fermions transform only under $SU(2)_R$ while the massive ones transform only under $SU(2)_L$.

\section{Regularization}\label{sec:reg}
Given a sum of loop integrals there are many ways to evaluate them. A standard approach for example is to evaluate the integrals using Feynman parameters and dimensional regularization. This can however be rather dangerous since we have powerlike divergences that introduce finite terms which may be difficult to determine unambiguously. In fact, as discussed in \cite{Roiban:2014cia}, this procedure would lead to a one-loop S-matrix not compatible with the symmetries of the BMN vacuum. Here we will therefore employ a similar regularization scheme as in \cite{Roiban:2014cia} which turned out to be compatible with the symmetries of the problem. It can be summarized as follows
\begin{enumerate}
\item Use the two algebraic identities
\bea
\frac{l_+l_-}{l^2-m^2}=&1+\frac{m^2}{l^2-m^2}\,,\label{eq:int-identity1}\\
\frac{1}{(l-P)^2-m^2}-\frac{1}{l^2-m^2}=&\frac{l_+P_-+l_-P_+-P^2}{(l^2-m^2)((l-P)^2-m^2)}\,,\label{eq:int-identity2}
\eea
on the integrand to reduce the divergent integrals to a minimal set of divergent integrals.
\item Use identities coming from allowing shifts of the loop variable to reduce them further (when consistent with the identities in the first step).
\item Evaluate the remaining integrals in dimensional regularization.
\end{enumerate}
The second step is not strictly necessary in our case since it is compatible with dimensional regularization. It is interesting to note that this procedure is quite powerful, for the NFS two-loop two-point function for example everything reduces to finite integrals after the first two steps. In addition one finds that the massless modes decouple from the calculations involving only massive external states.

\section{Two-point functions and dispersion relations}\label{sec:twopoint}
Here we will study the one- and two-loop two-point functions and corrections to the dispersion relation for massive and massless excitations. At one loop we will utilize the full BMN Lagrangian and compute the two-point function off-shell. The result contains UV-divergences which imply non-trivial wave function renormalization, just as in the case of pure RR-flux \cite{Roiban:2014cia}. This wave function renormalization will be important in the next section when relating the bare four-point function to the S-matrix. As we will see there is no one-loop correction to the dispersion relation so to find a correction one must go to two loops. We will do this in the NFS-limit, which simplifies the Lagrangian enough that we can carry out the computation.

\subsection{One-loop (off-shell) two-point functions}
Since there are no cubic interaction terms for the $AdS_3 \times S^3 \times T^4$ BMN-Lagrangian the only one-loop contributions to the two-point function comes from tadpole graphs
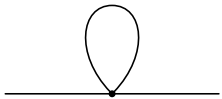
\bea\nn
\parbox[top][0.8in][c]{1.5in}{\fmfreuse{tadpole}}
\eea
The correction to the propagators of the massive bosons turn out to be UV-divergent. The divergence is proportional to $p^2-m^2$ and vanishes on-shell (the $p^2$-piece is due to the extra interaction terms in the Lagrangian compared to say $\phi^4$-theory where the tadpole just leads to mass renormalization) and has the interpretation of wave function renormalization (in the following we drop the tilde on all fields)
\bea 
\label{eq:wf-renormalization}
\langle zz \rangle = \frac{iZ_z}{p^2-m^2}+\mathcal{O}(g^{-2})\,,\qquad
\langle yy \rangle = \frac{iZ_y}{p^2-m^2}+\mathcal{O}(g^{-2})\,,
\eea
where the mass $m=\hat q$, interpolates between $m=1$ in the case of pure RR-flux and $m=0$ in the case of pure NSNS-flux. The wave function renormalization factors are given by
\bea 
Z_z=1+\frac{\hat q^2}{4\pi g}\gamma(\epsilon)\,, \qquad 
Z_y=1-\frac{\hat q^2}{4\pi g} \gamma(\epsilon)\,,
\label{eq:Zs}
\eea 
where the $\gamma(\epsilon)$ is the divergent part of a tadpole integral
\bea 
\int\frac{d^2k}{(2\pi)^2}\frac{1}{k^2-\hat q^2}\rightarrow\frac{i}{4\pi}\Big(-\frac{2}{\epsilon}+\gamma_E+\log\frac{\hat q^2}{4\pi}\Big)=
\frac{i}{4\pi}\gamma(\epsilon)\,,
\eea
in dimensional regularization in $2-\epsilon$ dimensions. As it turns out there is no one-loop correction to the two-point function for the massive fermions
\bea
\langle \chi^1\chi^1\rangle\vert_{1-loop}=0\,,\qquad\langle \chi^2\chi^2\rangle\vert_{1-loop}=0\,,
\eea
implying that the fermions are not subject to wave-function renormalization.

These results are essentially the same as the ones found for pure RR-flux $AdS_n\times S^n\times T^{10-2n}$ backgrounds in \cite{Roiban:2014cia}. The case of pure NSNS-flux is special since all excitations become massless and setting $\hat q=0$ we observe that the wave function renormalization disappears in this case. For the massless modes we find that the one-loop correction is zero off-shell for both bosons and fermions,
\bea
\langle u_Iu_I\rangle \vert_{1-loop} =0= \langle \chi^{r'}\chi^{r'}\rangle \vert_{1-loop}\,,
\eea 
from which we conclude that the massless modes do not get renormalized.

\subsection{Two-loop correction to dispersion relation in NFS limit}
As we have seen there is no one-loop correction to the dispersion relation and the first non-trivial correction arises at the two-loop level.\footnote{This is a simple consequence of the fact that we have no cubic interactions in $AdS_n\times S^n\times T^{10-2n}$ which means that the one-loop contribution comes from tadpole diagrams which cannot generate any non-trivial momentum dependence.} The diagrams that contribute to the two-point function at two loops are either of sunset type
\bea \label{diagram:sunset}
\parbox[top][0.8in][c]{1.5in}{\fmfreuse{sunset}}
\eea
or double tadpole type (there is also a six-vertex tadpole but it is subleading in the NFS-limit)
\bea
\parbox[top][0.8in][c]{1.5in}{\fmfreuse{doubblebubble}}
\eea
\\
The contribution from the latter diagram is actually zero. Therefore the non-trivial contribution comes only from the sunset diagrams. Due to the many contributions and the regularization issues involved this is very difficult to calculate utilizing the full BMN-Lagrangian. The problem is considerably simplified in the NFS limit where the left/right-moving sectors of the near-BMN string are boosted differently as $p_\pm \rightarrow g^{\mp1/2}p_\pm$. Keeping only the terms that are zeroth order in the coupling reduces the Lagrangian to (\ref{eq:LNFS}), with the quartic interactions given in (\ref{eq:L4NFS}). Nevertheless, the actual computation is still fairly complicated. A priori it involves both massive and massless modes propagating in the loops but surprisingly we will find that, just as was observed at one loop in \cite{Roiban:2014cia}, the massless modes decouple from the computation with massive external particles. This statement is true only in a specific regularization scheme however, which is similar to the one used at one loop.

In this section we will also present the corresponding computation for $AdS_5\times S^5$ and $AdS_2\times S^2\times T^6$, already performed in \cite{Klose:2007rz,Murugan:2012mf}. This is illuminating since it highlights the similarities and differences between the various $AdS\times S$-backgrounds.

The sunset diagram (\ref{diagram:sunset}) gives rise to integrals of the form
\bea
\label{eq:sunset-integrals}
I^{rs}_{[m_1m_2m_3]}=\int\frac{d^2k d^2l}{(2\pi)^4}\frac{k_-^r l_-^s}{(k^2-m_1^2)(l^2-m_2^2)((p-k-l)^2-m_3^2)}\qquad(r+s\leq4)
\eea 
where the internal particles are either all massive ($m=\hat q$ or $m=1$), one massive and two massless or all three massless. These integrals are power-counting UV divergent for $r+s>1$ and some care is required in evaluating them.\footnote{These integrals are actually finite in a Lorentz-invariant regularization scheme such as dimensional regularization. However, since the theory is not Lorentz-invariant in our case we have to worry about potential regularization ambiguities.} As explained in sec. \ref{sec:reg} our approach is to first use algebraic identities and then identities which follow from shifting the loop variables to reduce these to integrals with $r+s\leq1$ which are convergent and then evaluate these by Feynman parametrization. It is a non-trivial fact that this can actually be done since there are fewer identities than possible integrals that occur. The details of this reduction are given in appendix \ref{app:NFSids}. We will also comment on what happens if one evaluates all integrals directly in dimensional regularization using Feynman parameters. It turns out that the latter approach gives the same final answer but the decoupling of the massless modes is only seen in the first regularization scheme. Just as was the case a one loop we expect only the former approach to give an S-matrix consistent with the symmetries.

We will now consider each background in turn. We give the details for the bosonic modes although we have verified that the same goes through for the fermions as well.

\subsection*{$AdS_5\times S^5$}
This is the simplest case and the calculation was first done in \cite{Klose:2007rz}. Since all worldsheet excitations have the same mass $m=1$, only the integral in (\ref{eq:sunset-integrals}) with $m_1=m_2=m_3=1$ appears. The two-loop contribution to the two-point function for the bosons is given by
\begin{equation}
\label{eq:ads5ints}
\frac12
\Big(
-3I^{31}_{[111]}
-3I^{22}_{[111]}
+5p_-I^{21}_{[111]}
+3p_-^2I^{11}_{[111]}
+2p_-^2I^{20}_{[111]}\Big)\,.
\end{equation}
Using first the algebraic identities in eqs. (\ref{eq:alg1m})--(\ref{eq:alg2m}) to reduce this expression as far as possible and then the identities following from allowing shifts of the loop variable in eqs. (\ref{eq:shift1m})--(\ref{eq:shift4m}) to reduce the expression further one ends up with\footnote{In fact one obtains the same result using only the shift identities as was already seen in \cite{Klose:2007rz}.}
$$
-\frac{p_-^4}{3}I^{00}_{[111]}\,.
$$ 
This integral is finite and we get using (\ref{eq:i00-evaluated})
\bea
\varepsilon^2\vert_{2-loop}=-\frac{p_-^4}{192}\,,
\eea 
in agreement with the expansion of the exact dispersion relation (\ref{eq:disp-ads5}). Evaluating the integrals in (\ref{eq:ads5ints}) directly in dimensional regularization using Feynman parameters gives the same answer.

\subsection*{$AdS_3\times S^3\times T^4$}
For the case of $AdS_3\times S^3\times T^4$ with mixed flux we have two (complex) massive bosons and fermions, each with mass $\hat q$, and two (complex) massless bosons and fermions. The two-loop contribution to the two-point function for the massive bosons is given by\footnote{The action used here differs from the one in sec. \ref{sec:NFS} by some (unimportant) total derivative terms.}
\bea 
&& -\frac{1}{8}\hat q^4\Big( 4p_-^4 I^{00}_{[\hat q \hat q \hat q]}-13 p_-^3  I^{10}_{[\hat q \hat q \hat q]}+7 p_-^2(I^{11}_{[\hat q \hat q \hat q]}+I^{20}_{[\hat q \hat q \hat q]})
-p_-(  9 I^{21}_{[\hat q \hat q \hat q]}+2 I^{30}_{[\hat q \hat q \hat q]})+3 (I^{22}_{[\hat q \hat q \hat q]}+ I^{31}_{[\hat q \hat q \hat q]})\Big)
\nn\\
&& 
-\frac{1}{4}\hat q^4\Big(p_-^3  I^{01}_{[\hat q 00]}+p_-^2(2 I^{11}_{[\hat q 00]}+I^{02}_{[\hat q 00]})
-p_-(5I^{21}_{[\hat q 00]}+9I^{12}_{[\hat q 00]}+2I^{03}_{[\hat q 00]})
\label{eq:ads3ints}
\\
&&{}+2(3I^{22}_{[\hat q 00]}+2I^{13}_{[\hat q 00]}+I^{31}_{[\hat q 00]})\Big)\,.\nn
\eea 
Using first the algebraic identities in eqs. (\ref{eq:alg1})--(\ref{eq:alg2}), (\ref{eq:alg1m})--(\ref{eq:alg2m}) and the extra identities in eq. (\ref{eq:alg4m00}) and (\ref{eq:alg4mmm}) to reduce this expression as far as possible and then the identities following from allowing shifts of the loop variable in eqs. (\ref{eq:shift1})--(\ref{eq:shift4}) and (\ref{eq:shift1m})--(\ref{eq:shift4m}) to reduce the expression further one ends up with
\begin{equation}
\label{eq:ads3disp-massive}
-\hat q^4\frac{p_-^4}{3}I^{00}_{[\hat q \hat q \hat q]}=-\hat q^2\frac{p_-^4}{192}\,,
\end{equation}
where we again used (\ref{eq:i00-evaluated}). Note that the contribution from the massless modes coming from the integrals $I^{rs}_{[\hat q00]}$ in (\ref{eq:ads3ints}) has disappeared. Therefore, in this regularization scheme, the massless modes decouple from the two-loop two-point function of the massive modes.

Evaluating instead the integrals in (\ref{eq:ads3ints}) using dimensional regularization and Feynman parameters gives
\begin{equation}
\varepsilon^2\vert_{2-loop}=\hat q^2\Big(-\frac{\big(-6+\pi^2\big)p_-^4}{192\pi^2}\Big)_{[\hat q \hat q\hat q]}+\hat q^2\Big(-\frac{p_-^4}{32\pi^2}\Big)_{[\hat q00]}=-\hat q^2\frac{ p_-^4}{192}\,,
\end{equation}
where we have indicated which contribution comes from which type of integral. This gives the same final answer, which, as we will see below, agrees with the expansion of the exact dispersion relation (\ref{eq:disp-ads3}), but in this regularization scheme the massless modes clearly do not decouple.

For the massless bosonic modes the contribution to the two-loop two-point function is given by
\bea \nn
\langle u_1 u_1 \rangle\vert_{2-loop}:\quad -\frac{1}{2}\hat q^4 p_-^3 I^{10}_{[0\hat q\hat q]}, \qquad
\langle u_2 u_2 \rangle\vert_{2-loop}:\quad  -\frac{1}{2}\hat q^4 p_-^2\big(2 p_- I^{10}_{[0\hat q\hat q]}-2I^{11}_{[0\hat q\hat q]}-I^{20}_{[0\hat q\hat q]}\big)\,.
\eea 
Upon using the shift identity (\ref{eq:shift4}) and (\ref{eq:i10-evaluated}) both reduce to
\begin{equation}
\varepsilon^2\vert_{2-loop} = -\frac{1}{2}\hat q^4 p_-^3 I^{10}_{[0\hat q\hat q]} =  -\hat q^2 \frac{p_-^4}{32 \pi^2}\,,
\label{eq:ads3disp-massless}
\end{equation}
which differs by a factor of $6/\pi^2$ from that coming from (\ref{eq:disp-ads3}). It is worth noting that the contribution to the dispersion relation for the massless modes does involve the massive modes in the loops so the decoupling goes only in one direction. 

In appendix \ref{sec:massless-bmn} we go further and demonstrate that the two-point functions for massless modes are also (on-shell) UV-finite in the full BMN case.

\subsection*{$AdS_2\times S^2\times T^6$}
In the case of $AdS_2\times S^2\times T^6$ we have two real massive bosons and fermions with $m=1$ and three complex massless bosons and fermions. The contribution to the two-point function for the massive bosons is given by
\bea \nn 
&& \frac{1}{16}\Big(-p_-^3 I_{[111]}^{10}+p_-^2(-5I_{[111]}^{11}+3I_{[111]}^{20})+p_-(7I_{[111]}^{21}-6I_{[111]}^{30})-9(I_{[111]}^{22}+I_{[111]}^{31})\Big) \\ \nn
&& -\frac{3}{16}\Big(p_-^3 I_{[100]}^{01}+p_-^2(I_{[100]}^{11}+I_{[100]}^{02})-p_-(5 I_{[100]}^{21}+10I_{[100]}^{12}+2I_{[100]}^{03})+3(3I_{[100]}^{22}+I_{[100]}^{31}+2I_{[100]}^{13})\Big) \\ 
&& -\frac{3}{4}\Big(p_-I_{[000]}^{21}-I_{[000]}^{22}-I_{[000]}^{31}\Big)\,.
\label{eq:ads2ints}
\eea 
Using first the algebraic identities in eqs. (\ref{eq:alg1})--(\ref{eq:alg2}) and (\ref{eq:alg1m})--(\ref{eq:alg2m}) to reduce this expression as far as possible and then the identities following from allowing shifts of the loop variable in eqs. (\ref{eq:shift1})--(\ref{eq:shift4}) and (\ref{eq:shift1m})--(\ref{eq:shift4m}) to reduce the expression further one ends up with
\begin{equation}
-\frac14\frac{p_-^4}{3}I_{[111]}^{00}=-\frac{1}{4}\frac{p_-^4}{192}\,.
\end{equation}
We again note the decoupling of the massless modes just as in $AdS_3\times S^3\times T^4$. As was also seen at one loop in \cite{Roiban:2014cia} the ''effective'' coupling for $AdS_2\times S^2\times T^6$ differs by a factor of two compared to $AdS_3$ and $AdS_5$ giving rise to the extra factor of $\frac14$.

Evaluating instead the integrals in (\ref{eq:ads2ints}) using dimensional regularization and Feynman parameters gives (this calculation was done in \cite{Murugan:2012mf})
\begin{equation}
\varepsilon^2\vert_{2-loop}=\frac{1}{4}\Big(-\frac{\big(3+\pi^2\big)p_-^4}{192\pi^2}\Big)_{[111]}+\frac{1}{4}\Big(-\frac{3p_-^4}{64\pi^2}\Big)_{[100]}+\frac{1}{4}\Big(\frac{p_-^4}{32\pi^2}\Big)_{[000]}=-\frac{1}{4}\frac{p_-^4}{192}\,.
\end{equation}
Again we observe that in this regularization scheme the massless modes do not decouple.

For the massless bosonic modes the contribution to the two-loop two-point function is given by
\bea
\frac{1}{4}p_-^2\big(-p_-I_{[011]}^{10}+I_{[011]}^{11}\big)+\frac{1}{4}p_-^2\big(-p_-I_{[100]}^{01}+I_{[100]}^{11}+I_{[100]}^{02}\big)\,.
\eea
Using the algebraic identity (\ref{eq:alg1}), the shift identity (\ref{eq:shift4}) and (\ref{eq:i10-evaluated}) this reduces to
\begin{equation}
\varepsilon^2\vert_{2-loop}=-\frac{1}{4}p_-^3I_{[011]}^{10} =-\frac{1}{2}\frac{p_-^4}{32\pi^2}\,,
\end{equation}
where we, again, note that the contribution involves massive fields in the loops. It is interesting to note that here we find a relative factor of $1/2$ instead of $1/4$ as compared to the $AdS_3$ case.

\subsection{Comparison to proposed exact dispersion relation}
It is known that the two-loop corrections to the dispersion relation for the massive modes agree with the exact dispersion relation for $AdS_5 \times S^5$ and $AdS_2\times S^2\times T^6$ (with the additional factor of two accounted for)\footnote{In the latter case the dispersion relation is not completely fixed by the symmetries however \cite{Hoare:2014kma}.}. Here we will consider the case of $AdS_3\times S^3\times T^4$, which also contains the extra parameter $q$ measuring the amount of NSNS-flux. In other words we want to compare the correction in (\ref{eq:ads3disp-massive}) and (\ref{eq:ads3disp-massless}) with that coming from expanding the exact dispersion relation (\ref{eq:disp-ads3}) derived from the symmetries of the BMN vacuum. The latter takes the form \cite{Hoare:2013lja}
\begin{equation}
\varepsilon_\pm=\sqrt{M_\pm^2+4\hat q^2h^2\sin^2{\frac{\mathrm p}{2}}}\,,
\end{equation}
where the interpolating function appearing in the exact solution has an expansion
\begin{equation}
h=g+\frac{1}{g}h^{(2)}+\ldots\,.
\end{equation}
where $h^{(2)}$ denotes a possible two-loop correction (our one-loop calculation shows that there is no correction to $h$ at this order). It was recently argued in \cite{Lloyd:2014bsa} that the mass-like term should take the form\footnote{The dispersion relation of \cite{Hoare:2013lja} had $h$ in place of $g$.}
\begin{equation}
\label{eq:Mpm}
M_\pm=qg\mathrm p\pm\mathrm m\,,
\end{equation}
where $\mathrm m$ is $1$ for the massive excitations and $0$ for the massless ones. Performing the BMN-scaling of the momenta $\mathrm p=\frac{p}{h}$ and expanding at large $h$ we find
\begin{equation}
\varepsilon_\pm^2=
\mathrm m^2\hat q^2
+(p\pm\mathrm mq)^2
-2qp(qp\pm\mathrm m)\frac{h^{(2)}}{h^2}
-\hat q^2\frac{p^4}{12h^2}
+\mathcal O(h^{-3})
\end{equation}

In order to compare with the NFS computation eq. (\ref{eq:ads3disp-massive}) and eq. (\ref{eq:ads3disp-massless}) we need to go to the form of the dispersion relation that is relativistic at lowest order since this is the case for the form of the quadratic action we are using. To do this we write $\tilde p=p\pm q$. Performing the NFS scaling $\tilde p_\pm\rightarrow h^{\mp 1/2}\tilde p_\pm$ gives\footnote{In our conventions, $p=\frac{1}{2}(p_+-p_-)$.}
\begin{equation}
\label{eq:NFS-interpolating-disp}
\varepsilon^2=\mathrm m^2\hat q^2+\tilde p^2-\hat q^2\frac{\tilde p_-^4}{192}
+\mathcal O(h^{-1/2})\,,
\end{equation}
in nice agreement with our result from worldsheet perturbation theory for the massive modes (\ref{eq:ads3disp-massive}). An earlier suggestion \cite{Hoare:2013ida} for the form of the dispersion relation involving $\sin{\frac{\mathrm p}{2}}$ instead of the term linear in $\mathrm p$ in (\ref{eq:Mpm}) is ruled out as it would lead to a coefficient of $1$ instead of $\hat q^2$ for the $\tilde p_-^4$-term. Our calculation is not sensitive to the difference between $h$ and $g$ in (\ref{eq:Mpm}), however. This calculation shows that the two-loop correction to the massless dispersion relation should be the same as the correction to the massive one. This disagrees with our result (\ref{eq:ads3disp-massless}) by a factor of $6/\pi^2$ as discussed in sec. \ref{sec:intro}.

\section{S-matrix at one loop}\label{sec:Smatrix}
The tree-level S-matrix for two-to-two scattering of massive modes was found in \cite{Hoare:2013pma}. Here we will extend the analysis to one loop and compare to the conjectured form of the all loop S-matrix \cite{Hoare:2013ida} including dressing phases \cite{Engelund:2013fja,Babichenko:2014yaa,Bianchi:2014rfa}. Since the analysis is very similar to the case of zero NSNS-flux described in \cite{Roiban:2014cia}, we will be rather brief in technical detail in this section.  For simplicity we will furthermore restrict to purely bosonic in and out-states. For scattering processes involving fermions we need the quartic fermion terms in the Lagrangian which, although explicitly known, are quite involved in the case of mixed flux. 

The two-body S-matrix takes the form,
\bea
\mathbbm{S}=\mathbbm{1}+i \mathbbm{T}, \qquad 
\mathbbm{T}=\frac{1}{g}\mathbbm{T}^{(0)}+\frac{1}{g^2}\mathbbm{T}^{(1)}+\dots
\eea
where the superscript denotes the loop order. The $\mathbbm{T}$-matrix maps a two-particle in-state to a two-particle out-state as,
\bea
\mathbbm{T}| A(p) B(k) \rangle = T^{CD}_{AB}|C(p) D(k)\rangle\,.
\eea 
At the one-loop order the relevant diagrams are four-vertex s,t and u-channel diagrams
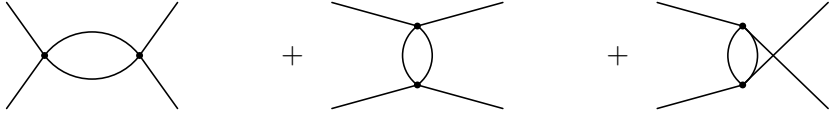
\bea\nn
\parbox[top][0.8in][c]{1.5in}{\fmfreuse{schannel}} + \parbox[top][0.8in][c]{1.5in}{\fmfreuse{tchannel}}+\parbox[top][0.8in][c]{1.5in}{\fmfreuse{uchannel}}
\eea
and the six-vertex tadpole diagram
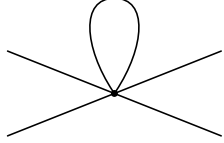
\bea \nn
\parbox[top][0.8in][c]{1.5in}{\fmfreuse{tadpolesix}}
\eea
These give rise to standard bubble and tadpole-type integrals of the form\footnote{The relevant four-vertices are such that the virtual particles in the bubble diagrams always have the same mass.}
\bea
B^{rs}(P)=\int \frac{d^2l}{(2\pi)^2}\frac{l_+^r l_-^s}{(l^2-m^2)((l-P)^2-m^2)}, \qquad T^{rs}(P)=\int \frac{d^2l}{(2\pi)^2}\frac{l_+^r l_-^s}{(l-P)^2-m^2}\,,
\label{eq:one-loop-int}
\eea
where $P$ is a linear combination of the external momenta $p$ and $k$ and $m$ is either $\hat q$ or $0$. Many of these integrals are UV-divergent and require regularization. As described in \cite{Roiban:2014cia}, using only the two identities (\ref{eq:int-identity1}) and (\ref{eq:int-identity2}) on the bubble integrands and allowing shifts of the loop variable in the tadpole integrals we can reduce all integrals to the following three
\bea
B^{00}(P)\,,\qquad T^{00}(0)\,,\qquad T^{11}(0)\,.
\eea 
The last two are divergent and we evaluate them in dimensional regularization.

\subsection{Scattering amplitudes and dressing phases}
We start with considering scattering of $z$ and $y$ bosons separately,
\bea 
&&
\mathbbm{T} \ket{z_\pm z_\pm} = \ell_1^z \ket{z_\pm z_\pm}, \qquad 
\mathbbm{T} \ket{y_\pm y_\pm} = \ell_1^y\ket{y_\pm y_\pm}, \\ \nn
&& 
\mathbbm{T} \ket{z_\pm z_\mp} = \ell_2^z \ket{z_\pm z_\mp}, \qquad 
\mathbbm{T} \ket{y_\pm y_\mp} = \ell_2^y \ket{y_\pm y_\mp} 
\eea 
where each amplitude is given, {\it before} taking into account the wave function renormalization eqs. (\ref{eq:wf-renormalization}) and (\ref{eq:Zs}), by
\bea 
&& \ell_1^z=\frac{1}{g}l_1^z + \frac{1}{g^2}\Big( 2 \Theta_{\pm\pm}^{LL}-\frac{1}{2\pi}\gamma(\epsilon) l_1^z\Big)\,, \qquad 
\ell_1^y=\frac{1}{g}l_1^y + \frac{1}{g^2}\Big( 2 \Theta_{\pm\pm}^{LL}+\frac{1}{2\pi}\gamma(\epsilon) l_1^y\Big)\,, \\ \nn
&&
\ell_2^z=\frac{1}{g}l_2^z + \frac{1}{g^2}\Big( 2 \Theta_{\pm\mp}^{LR}-\frac{1}{2\pi}\gamma(\epsilon) l_2^z\Big)\,, \qquad 
\ell_2^y=\frac{1}{g}l_2^y + \frac{1}{g^2}\Big( 2 \Theta_{\pm\mp}^{LR}+\frac{1}{2\pi}\gamma(\epsilon) l_2^y\Big)\,.
\eea 
Here we introduced the notation $\Theta_{\pm\pm}^{LL}$ and $ \Theta_{\pm\pm}^{LR}$ for the one-loop phases of \cite{Engelund:2013fja,Bianchi:2014rfa,Babichenko:2014yaa}. The tree-level expressions $l_1,\,l_2$ can be found in \cite{Hoare:2013pma}. Note that we have dropped the imaginary terms in the amplitude since these are completely determined in terms of tree-level amplitudes via the optical theorem and can therefore be trivially reinstated.

For the scattering of bosons with different flavors we find
\bea
\mathbbm{T} \ket{z_\pm y_\pm} = \ell_3 \ket{z_\pm y_\pm}, \qquad 
\mathbbm{T} \ket{z_\pm y_\mp} = \ell_4\ket{z_\pm y_\mp}, 
\eea
with
\bea
\ell_3=\frac{1}{g}l_3 + \frac{1}{g^2}2\Theta_{\pm\pm}^{LL}\,,\qquad\ell_4=\frac{1}{g}l_4 + \frac{1}{g^2}2\Theta_{\pm\mp}^{LR}\,.
\eea
It is easy to see that renormalizing the bosons according to eqs. (\ref{eq:wf-renormalization}) and (\ref{eq:Zs}) renders all amplitudes finite. Furthermore, since the renormalization of $z$ and $y$-particles come with opposite signs it is clear why the above mixing processes are finite prior to any renormalization. This furthermore implies that the processes
\bea \nn
 \mathbbm{T} \ket{y_\pm y_\mp} = \ell_5 \ket{z_\pm z_\mp }+\dots
\eea 
should be finite. A quick calculation indeed shows that this is the case.

Let us now look at the explicit expression for the phases. In order to make connection with known results from the literature we should undo the rotation in (\ref{eq:quadratic-shift}). That is, we should shift the worldsheet momenta as $p \rightarrow p \mp q$. Doing this we find the following compact expressions for the phase-factors:
\bea
\label{eq:LR-phase}
&& \Theta^{LL}_{\pm \pm}(p,k)= \\ \nn
&& \frac{p^2 k^2\Big(\varepsilon_\pm(p)\varepsilon_\pm(k)+(p\pm q)(k\pm q)+\hat q^2\Big)}{2\pi(p-k)^2}\log\frac{p_-}{k_-}+\frac{p \,k (p+k)\big(\varepsilon_\pm(p) k + \varepsilon_\pm(k) p \big)}{4\pi(p-k)}, \\ \nn
&& \Theta^{LR}_{\pm \mp}(p,k)= \\ \nn
&& \frac{p^2 k^2\Big(\varepsilon_\pm(p)\varepsilon_\mp(k)+(p\pm q)(k\mp q)-\hat q^2\Big)}{2\pi(p+k)^2}\log\frac{p_-}{k_-} -\frac{p \,k (p-k)\big(\varepsilon_\pm(p) k + \varepsilon_\mp(k) p \big)}{4\pi(p+k)},  
\eea
where $\varepsilon_\pm=\sqrt{\hat q^2+(p\pm q)^2}$. The above phases exactly match the proposals in \cite{Engelund:2013fja,Bianchi:2014rfa,Babichenko:2014yaa}. 

As a final comment we would also like to mention that in the regularization scheme employed here, the massless modes completely decouple. That is, simply putting all massless fields to zero in the Lagrangian, which is the same as using the supercoset sigma model, gives the same one-loop massive S-matrix.

\section{Conclusions}\label{sec:conclusions}

In this paper we have performed one- and two-loop calculations on the string worldsheet. At one loop the analysis was a continuation of the work initiated in \cite{Roiban:2014cia}, where it was shown that the one-loop massive sector BMN S-matrix is finite and agrees with the form expected from symmetries and crossing once wave function renormalization and regularization issues are accounted for. For the case at hand, with mixed NSNS- and RR-fluxes, the situation is similar including the same form of wave function renormalization for the massive bosons and, with the proper regularization scheme, perturbative computations are in complete agreement with the proposal of \cite{Hoare:2013pma} based on symmetries. Just as was the case in \cite{Roiban:2014cia} we again find that in this regularization the massless modes decouple from the calculation of the massive sector S-matrix.

We also performed a two-loop calculation of the correction to the two-point function of massive and massless modes. Due to the somewhat involved sunset diagrams appearing this part of the analysis was performed in the simpler Near-Flat-Space limit. Somewhat surprisingly we again found that the massless modes decouple from the calculations in the massive sector (again in the proper regularization scheme). This property is not symmetric however and for massless modes the massive modes do contribute to the dispersion relation. In the case of massive modes the two-loop correction exactly matches the one coming from (the NFS-limit of) the exact dispersion relation (\ref{eq:disp-ads3}). For the massless modes however the correction does not seem to agree with the proposed exact dispersion relation \cite{Lloyd:2014bsa,Hoare:2014kma}. While the correction still involves the fourth power in momentum, it differs by a factor of $6/\pi^2$, regardless of the amount of NSNS-flux $q$. The origin of this mismatch is rather mysterious and currently it is not clear to us how to reconcile our result with the proposed exact dispersion relation for the massless modes. Perhaps the central charges (\ref{eq:centralcharges}) receive additional quantum corrections or maybe the massless asymptotic states should be changed in comparing the two cases. 

This work opens up the way for several future lines of research. Perhaps the most pressing question is to reconcile the perturbative computations of the massless dispersion relation with the proposed exact dispersion relation. As a first step one would like to compute the correction to the dispersion relation in the full BMN case rather than the truncated NFS-limit, which does not see lower powers of the momenta in the dispersion relation. Furthermore, computing one- and two-loop S-matrix elements involving massless coordinates would allow us to better understand the reason for the mismatch. While the phases for the massless sector are currently not known, perhaps a comparison can be made with the results in \cite{Engelund:2013fja,Bianchi:2014rfa,Babichenko:2014yaa}. It would also be interesting to perform the two-loop computations for massive modes in the full BMN setting. This would for example show if there is a two-loop correction to the interpolating function $h$. However, these computations are fairly complicated and have, as of yet, not even been performed in the simplest case of $AdS_5\times S^5$. 

We plan to address some of these questions in the near future. 

\section*{Acknowledgments}
We wish to thank R. Roiban and A. Tseytlin for collaboration at earlier stages in this project and for many helpful discussions and comments. We also thank S. Penati and O. Ohlsson Sax for useful discussions and comments. The work of PS was supported by a joint INFN and Milano-Bicocca postdoctoral grant. The  work of LW was supported by the ERC Advanced grant No.290456 ``Gauge theory -- string theory duality''.

\appendix
\section*{Appendix}
\section{Integral identities for NFS sunset integrals}\label{app:NFSids}
The NFS sunset integrals are of the form \ref{eq:sunset-integrals}. Since only integrals with at least two masses equal occur we can label the loop momenta so that $m_3=m_2$. Using the identity (\ref{eq:int-identity2}) with $m=m_2$ and $P=p-k$ and multiplying with $k_-^{r+1}l_-^{s+1}$ we derive, after a bit of algebra, the identity
\begin{align}
&p_+(I^{r+1,s+2}+I^{r+2,s+1})
+m_1^2(p_-I^{r,s+1}-I^{r,s+2}-I^{r+1,s+1})
+m_2^2(p_-I^{r+1,s}-I^{r+2,s})
-p^2I^{r+1,s+1}
\nonumber\\
=
&I^{r,s+2}_{-1}
+I^{r+1,s+1}_{-1}
-p_-I^{r,s+1}_{-1}
+I^{r+2,s}_{-2}
+I^{r+1,s+1}_{-2}
-p_-I^{r+1,s}_{-2}
-I^{r+1,s+1}_{-3}\,,
\label{eq:alg-id-NFS}
\end{align}
where $I^{rs}=I^{rs}_{[m_1m_2m_2]}$ and the tadpole-like integrals $I^{rs}_{-1}$, $I^{rs}_{-2}$, $I^{rs}_{-3}$ are the same as $I^{rs}$ but with the first, second and third factor in the denominator dropped respectively. We will only need the cases $(r,s)=\{(0,0),\,(1,0),\,(0,1)\}$ which read ($p$ denotes the external momentum)
\begin{align}
p_+(I^{12}+I^{21})+m_1^2(p_-I^{01}-I^{02})+m_2^2(p_-I^{10}-I^{20})-(p^2+m_1^2)I^{11}=&0\,,\label{eq:alg1}\\
p_+(I^{22}+I^{31})+m_1^2(p_-I^{11}-I^{12})+m_2^2(p_-I^{20}-I^{30})-(p^2+m_1^2)I^{21}=&0\,,\label{eq:alg2}\\
p_+(I^{13}+I^{22})+m_1^2(p_-I^{02}-I^{03})+m_2^2(p_-I^{11}-I^{21})-(p^2+m_1^2)I^{12}=&0\,.\label{eq:alg3}
\end{align}
Here we have assumed that shifts of loop variables are allowed in the tadpole integrals, which then turn out to cancel out in the RHS of (\ref{eq:alg-id-NFS}). Note that these algebraic identities are typically not consistent with dimensional regularization (this is why our regularization scheme is different).

Allowing shifts such as $l\rightarrow p-l-k$ in the loop variables in $I^{rs}_{[m_1m_2m_2]}$ leads to the additional identities (we list only the ones relevant for us)
\begin{align}
2I^{21}+I^{30}-p_-I^{20}&=0\,,\label{eq:shift1}\\
2I^{13}+I^{31}+3I^{22}-2p_-I^{21}-3p_-I^{12}+p_-^2I^{11}&=0\,,\label{eq:shift2}\\
4I^{03}+2I^{21}+6I^{12}-4p_-I^{11}-6p_-I^{02}+p_-^2I^{10}+4p_-^2I^{01}-p_-^3I^{00}&=0\,,\label{eq:shift3}\\
2I^{11}+I^{20}-p_-I^{10}&=0\,.\label{eq:shift4}
%
\end{align}
These identities are compatible with dimensional regularization since shifts of the loop momenta are allowed in that regularization scheme.

In the special case when all masses are equal, i.e. $m_1=m_2=m$, $I^{rs}$ is symmetric in $(r,s)$ and the identities simplify to (here $I^{rs}=I^{rs}_{[mmm]}$)
\begin{align}
2p_+I^{21}+2m^2(p_-I^{10}-I^{20})-(p^2+m^2)I^{11}=&0\,,\label{eq:alg1m}\\
p_+(I^{31}+I^{22})+m^2(p_-I^{11}+p_-I^{20}-I^{30})-(p^2+2m^2)I^{21}=&0\,,\label{eq:alg2m}
\end{align}
from algebraic manipulations eqs. (\ref{eq:alg1})--(\ref{eq:alg3}) and
\begin{align}
3I^{31}+3I^{22}-5p_-I^{21}+p_-^2I^{11}&=0\,,\qquad 2I^{21}+I^{30}-p_-I^{20}=0\,,\label{eq:shift1m}\\
2I^{11}+I^{20}-p_-I^{10}&=0\,,\qquad 3I^{10}-p_-I^{00}=0\,,\label{eq:shift4m}
\end{align}
from shifting the loop variables eqs. (\ref{eq:shift1})--(\ref{eq:shift4}).

It turns out that these identities are all that we need except for in the case of the correction to the massive dispersion relation in $AdS_3\times S^3\times T^4$. In that case we are left with terms involving $I^{20}$ and $I^{02}$ which can not be reduced further with the above identities. It turns out however that we can derive more algebraic identities by using identities for the one-loop bubble integrals inside these two-loop integrals. In \cite{Roiban:2014cia} we showed how, by using the algebraic identity (\ref{eq:int-identity2}), one can derive, among others, the identity
\begin{equation}
B^{02}(P)-\frac{P_-}{P_+}(\frac12P^2-m_2^2)B^{00}(P)=0\,,
\label{eq:B02id}
\end{equation}
where $B^{rs}(P)$ are the one-loop bubble integrals defined in (\ref{eq:one-loop-int}) with mass $m=m_2$. Taking $P=p-k$, multiplying with $\frac{1}{k^2-m_1^2}$ and integrating over $k$ we find
\begin{equation}
\label{eq:appendix-identity}
2I^{02}
-I^{20}
+2p_-I^{10}
-p_-^2I^{00}
+2m_2^2\tilde I
=0\,,
\end{equation}
where
\begin{equation}
\tilde I=\int\frac{d^2kd^2l}{(2\pi)^4}\frac{(p_--k_-)^2}{(p-k)^2(k^2-m_1^2)(l^2-m_2^2)((p-k-l)^2-m_2^2)}\,.
\end{equation}
Taking $m_2=0$ we find the identity we need in that case
\begin{equation}
2I^{02}_{[m00]}-I^{20}_{[m00]}+2p_-I^{10}_{[m00]}-p_-^2I^{00}_{[m00]}=0\,.
\label{eq:alg4m00}
\end{equation}
To get an identity for non-zero $m_2$ we must evaluate the integral $\tilde I$. We can do this any way we like since it is UV and IR finite. A more efficient way than trying to do it directly is to use again algebraic and shift identities. From the algebraic identity
\begin{equation}
\frac{1}{k^2-m_1^2}-\frac{1}{(p-k)^2}=\frac{m_1^2-p^2+p_+(p_--k_-)+p_-(p_+-k_+)}{(p-k)^2(k^2-m_1^2)}\,,
\end{equation}
we see that if $p^2=m_1^2$ we can multiply with $\frac{p_--k_-}{(l^2-m_2^2)((p-k-l)^2-m_2^2)}$ and integrate over $k$ and $l$ to get
\begin{equation}
0=
p_+\tilde I
+I^{10}
+\int\frac{d^2kd^2l}{(2\pi)^4}\frac{p_--k_-}{(p-k)^2(l^2-m_2^2)((p-k-l)^2-m_2^2)}
=
p_+\tilde I
+I^{10}
-I^{10}|_{m_1,p=0}\,,
\end{equation}
where in the last step we shifted $k\rightarrow p+k$. However, 
\bea \nn
I^{10}|_{m_1,p=0}=I^{10}_{0mm}\vert_{p = 0}\,,
\eea
and since the integral is proportional to $p_-$ it vanishes at $p=0$ and we have,
\bea
\nn 
p_+\tilde I + I^{10}=0\,.
\eea 
Thus, for $p^2=m_2^2=m_1^2=m^2$, we finally find, using the above and equation (\ref{eq:appendix-identity}),
\begin{equation}
I^{20}_{[mmm]}-p_-^2I^{00}_{[mmm]}=0\,,
\label{eq:alg4mmm}
\end{equation}
which is the extra identity we need in the case of $AdS_3\times S^3\times T^4$.

\subsection{Remaining integrals}
After the algebraic and shift identities in the above section have been employed we end up with two remaining integrals, $I^{00}_{[mmm]}$ and $I^{10}_{[0mm]}$, that we need to evaluate. Since the integrals are finite we can evaluate them as we like. Using Feynman parameters we have, for the massive integral, with $ p^2=m^2$,
\bea
\label{eq:i00-evaluated}
I^{00}_{[mmm]} =  \frac{1}{16\pi^2 m^2}\int_0^1 dx_1 dx_2 dx_3 \frac{\delta(1-x_1-x_2-x_3)}{(x_1+x_2)(x_1+x_3)(x_2+x_3)}=\frac{1}{64 m^2}\,,
\eea
while for the massless one, with $p^2=0$, we find
\bea
\label{eq:i10-evaluated}
I^{10}_{[0mm]} =  \frac{p_-}{16\pi^2 m^2}\int_0^1 dx_1 dx_2 dx_3 \frac{\delta(1-x_1-x_2-x_3)x_2 x_3 }{(x_2+x_3)(x_1x_2 + x_1 x_3+x_2x_3)^2}=\frac{p_-}{16 \pi^2 m^2}\,.
\eea

\section{Two-loop finiteness of massless BMN two-point function}
\label{sec:massless-bmn}
Since we find a result for the massless modes which does not have the form proposed in \cite{Lloyd:2014bsa} let us demonstrate that the (on-shell) two-point function is finite in the full BMN case. This is important since it demonstrates the quantum consistency of the theory, yielding validity to our result for the two-loop correction to the dispersion relation in (\ref{eq:nfs-dispersion-relation}) and suggesting that it is not simply a regularization artifact. The actual computation is rather complicated and involves heavy use of the identities (\ref{eq:int-identity1}) and (\ref{eq:int-identity2}) and we will postpone the technical details to a forthcoming paper (where we will also calculate the finite part of the two-point function). 

When we go beyond the strict NFS-limit there are three divergent classes of diagrams and as before the most complicated to evaluate are the sunset-diagrams. Using (\ref{eq:int-identity1}) and (\ref{eq:int-identity2}) repeatedly we find,
\bea
\parbox[top][0.8in][c]{1.1in}{\fmfreuse{sunset}}=-\frac{1}{4}\gamma(\epsilon)^2 \big(p_-^2+p_+^2\big)+\textrm{finite terms}.
\eea 
The second class of diagrams to evaluate is the double tadpole built out of four-vertices. A simpler calculation gives,
\bea
\parbox[top][0.8in][c]{1.5in}{\fmfreuse{doubblebubble}}=\frac{1}{4}\gamma(\epsilon)^2 \big(p_-^2+p_+^2\big)\,,
\eea\\
which precisely cancels the divergent contribution from the sunset-diagram. Thus, if we can show that the double tadpole built out of the six-vertex Lagrangian is zero we are done. Performing the computation shows that this amplitude is proportional to $\gamma(\epsilon)^2p_+p_-$ which, while non-zero off-shell, vanishes once we put the external legs on-shell.\footnote{Note that we have not determined the numerical coefficient in front of the six-vertex tadpole divergence. One contribution comes from the $F^4B^2$ terms of the Lagrangian which are rather complicated so we have not written them down explicitly. However, it easy to see that the relevant quartic fermion terms are multiplied with two massless bosons of the form $\partial_\pm \bar u \partial_\mp u$. These types of terms can only give a contribution like $p_+ p_-$ while the two internal fermion tadpole-loops gives something proportional to $\gamma(\epsilon)^2$ and thus the final result is finite on-shell.}

\bibliographystyle{JHEP}
\bibliography{ads2ads3etc}
\end{document}